%% file: ms_v6.tex
\documentclass[usenatbib]{mn2e}
\usepackage{fixltx2e}
\usepackage{mathptmx}
\usepackage{color}
\usepackage{graphics,graphicx,amssymb,amsmath,fixltx2e,natbib,epsfig}
\usepackage{amsmath}

\def\gtrsim{\lower 2pt \hbox{$\, \buildrel {\scriptstyle >}\over
{\scriptstyle \sim}\,$}}
\def\lesssim{\lower 2pt \hbox{$\, \buildrel {\scriptstyle <}\over
{\scriptstyle \sim}\,$}}

\def\xmm{{\sl XMM-Newton}}

\def\chandra{{\sl Chandra}}

\begin{document}

\title{{\sl Chandra} survey of nearby highly inclined disk galaxies -- IV: 
New insights into the working of stellar feedback}
\author[]{Q. Daniel Wang$^{1,2}$\thanks{E-mail:wqd@astro.umass.edu}, Jiangtao Li$^{3}$, Xiaochuan Jiang$^{4}$,  and Taotao Fang$^{4}$\\ 
$^{1}$Department of Astronomy, University of Massachusetts, 710 North Pleasant St., Amherst, MA, 01003, U.S.A.\\
$^{2}$School of Astronomy and Space Science, Nanjing University, Nanjing 210093, China\\
$^{3}$Department of Astronomy, University of Michigan, 311 West Hall, 1085 S. University Ave, Ann Arbor, MI, 48109-1107, U.S.A.\\
$^{4}$Department of Astronomy and Institute for Theoretical Physics and Astrophysics, Xiamen University, 422 Siming South Road, Siming, Xiamen, Fujian, China}



\maketitle

\label{firstpage}

\begin{abstract}
Galaxy evolution is regulated by the interplay between galactic disks and their surrounding medium. We study this interplay by examining how the galactic coronal emission efficiency of stellar feedback depends on the (surface and specific) star formation rates (SFRs) and other parameters for a sample of 52 \chandra-observed nearby highly inclined disk galaxies. We first measure the star forming galactic disk sizes, as well as the SFRs of these  galaxies, using data from the Wide-Field Infrared Survey Explorer, and then show that 1) the specific 0.5-2~keV luminosity of the coronal emission correlates with the specific SFR in a {\sl sub-linear} fashion:  on average, $L_X/L_K \propto (SFR/M_*)^{\Gamma}$ with $\Gamma =0.29\pm0.12$; 2) the efficiency of the emission  $ L_X/SFR$ decreases with increasing surface SFR ($I_{SFR}$; $\Gamma = -0.44\pm0.12$); and 3) the characteristic temperature of the X-ray-emitting plasma weakly correlates with $I_{SFR}$ ($\Gamma = 0.08\pm0.04$). These results, somewhat surprising and anti-intuitive, suggest that a) the linear correlation between $L_X$ and SFR, as commonly presented, is largely due to the correlation of these two parameters with galaxy mass; b) much of the mechanical energy from stellar feedback likely drives global outflows with little X-ray cooling and with a mass-loading efficiency  decreasing fast with increasing $I_{SFR}$ ($\Gamma \lesssim -0.5$); c) these outflows heat and inflate the medium around the galactic disks of massive galaxies, reducing its radiative cooling rate, whereas for relatively low-mass galaxies, the  energy in the outflows is probably dissipated in regions far away from the galactic disks.
\end{abstract}


\section{Introduction}

Disk galaxies are complex ecosystems --- manifestations of the interplay among such processes as accretion from the intergalactic medium (IGM), star formation (SF), and the feedback from stars and active galactic nuclei (e.g., ~\citealt{White91,Bower12,Hopkins14,Keller15,Mitra15,Agertz15}). However, how the interplay actually works remains poorly understood~\citep[e.g.,][]{Lu14}.  A potentially powerful tool for probing the interplay is observations of its expected product --- galactic coronae, or diffuse hot gas around the disks. 

Indeed, galactic coronae have commonly been observed with modern X-ray 
observatories (mostly \chandra\ and \xmm), especially around nearby 
edge-on disk galaxies, for which the contamination from point-like sources is minimal (e.g., ~\citealt{Wang00,Strickland04a,Tullmann06,Li08,Wang10,Li13a, Li13b}, hereafter, Papers~I and II). The observed X-ray emission shows signatures of optically-thin thermal origin, although contributions from other related processes such as charge exchange (or transfer) at the interface between hot and cool gases can also be significant (e.g., ~\citealt{Liu11,Liu12,Zhang14}). In any case, the emission seems to be a good tracer of the coronae and/or their interaction with cool gas (e.g., ~\citealt{Strickland02}).

What the coronae exactly represent is still not clear. They could be due largely to the heating of the stellar feedback-driven galactic fountains/gravitationally unbound winds, inflows from the IGM accretion, or some mixture of these components. Also uncertain are the volume-filling factor of the coronal gas and hence its mass and energy contents, as well as the astrophysics involved in the heating/cooling and at the interface between hot and cool gas phases (e.g., ~\citealt{Strickland09,Lagos13}). The relative importance of these phenomena and processes may vary strongly from one region to another in a galaxy and may depend on its circumstance (mass, SFR, environment, etc.). These uncertainties thus present a major problem to the development of a comprehensive theory of  the galaxy formation and evolution~\citep[e.g.,][]{Lu14}. A sensible way to begin to address this problem is to empirically characterize the dependencies of the X-ray emission on key galaxy parameters (a kind of study similar to 
finding the SF law). 

Such empirical studies have surfaced in recent years. It has been shown that the coronal luminosity ($ L_X$) strongly 
correlates with the star formation rate (SFR; e.g., ~\citealt{Strickland04b,Grimes05,Tullmann06,Li08}; Paper II). This correlation is
commonly characterized with a fitted linear relation of 
$L_X$ vs. SFR (i.e., assuming a power law slope $\Gamma =1$). However, the dispersion around this relation is large, which can be measured by the root mean 
square of the data around it 
(RMS$\approx 0.52$ dex; Paper II), and is not well understood. 

In Papers~I and II, we have systematically studied the galactic coronae associated with a sample of 53 nearby highly-inclined disk galaxies using \emph{Chandra} observations. We find a tighter correlation between $ L_X$ and the supernova (SN) energy input rate ($\dot{E}_{SN}$, RMS $\approx 0.48$ 
dex). The metal abundance ratio of the coronal gas depends on galaxy type
and is consistent with the expected feedback mainly from core collapsed 
and/or Type~Ia SNe. These results indicate
that the X-ray emission traces SN feedback. We also show that the X-ray 
radiation efficiency ($\eta = L_X/\dot{E}_{SN}$) 
of the coronae is only $\sim 0.4\%$, on average, indicating that the feedback
energy is consumed primarily in other forms, which remain uncertain. 
There are signs for potential dependencies of $L_X$ on other galaxy 
parameters such
as the gravitational mass (as traced by the rotation velocity of a galaxy).
But most of these signs are very weak: examples are $\eta $ vs. $SFR$ with a Spearman's rank order correlation coefficient $r_s=-0.22\pm0.17$, 
or $\eta $ vs. the surface core-collapsed SN rate with $r_s=-0.38\pm0.14$
for our entire sample of the galaxies (Paper II; more discussion later). The X-ray 
measurements are further compared to cosmological hydrodynamical simulations of galactic coronae 
in \citet{Li14} (Paper~III). In addition to highlighting various limitations
of these simulations, this comparison shows that the specific 
X-ray emission (e.g., $L_X/M_*$) is enhanced for most massive disk galaxies
($M_* \gtrsim 2 \times 10^{11} M_\odot$), presenting a signature 
for the X-ray contribution from the IGM accretion 
(see also \citealt{Anderson11,Anderson16,Dai12,Bogdan13,Bogdan15}).

Here we report results from a re-examination of the correlations of $L_X$ with key galaxy properties with multiple improvements. While the same sample of \chandra-observed galaxies as analyzed in Paper I and II are used here,  we find that various correlations can be significantly tightened when galaxy-distance independent parameter ratios of the galaxies (e.g., $L_X/L_{K}$ and SFR/$M_{*}$) are 
adopted. The use of the parameter ratios let us avoid the correlation effect due to the uncertainty in 
galaxy distances. There is an uncomfortable range of distances in the literature for some of the galaxies; it is not uncommon to find a factor $\gtrsim 2$ difference  (e.g., ~\citealt{Wiegert15}). We study 
the nature of the  linear correlation between $ L_X$ and SFR: Is 
it because both of these two parameters are correlated with galaxy mass (e.g.,
Paper II; \citealt{Brinchmann04,Mitchell14})? 
Furthermore, we use  the recently released data 
from the Wide-field Infrared Survey Explorer ({\sl WISE}) to estimate both the rate and the disk area of the SF in individual galaxies, which enables us for the first time to examine the dependence of $L_X$ on the surface SFR ($I_{SFR}$). 
We aim to constrain the mass and energetics of galactic outflows, as well as their interplay with
the surrounding medium. 

The rest of the paper is organized as follows: In \S~\ref{s:obs}, we describe 
the data that are used in the present analysis. We present the analysis and
results in \S~\ref{s:res} and discuss their implications  in \S~\ref{s:dis}.
Finally in \S~\ref{s:sum}, we give a summary of the work and point out future directions
of the research.

\section{Description of the data and reduction}\label{s:obs}
We estimate the SFRs and SF disk sizes from the {\it WISE} 22~\micron\ (W4 band) images of our sample galaxies~\citep{Wright10,jarrett2013}. The emission
in this band is sensitive to warm dust and hence to its primary 
heating source, UV radiation from massive stars. The data reduction and analysis of
the {\it WISE} data in the band, as well as in the others, will be detailed in Jiang et al. (in preparation). But briefly, first we approximate the outer boundary of a galaxy as an ellipse defined with its center coordinates, apparent major/minor axes, and  position angle (measured in the 2MASS ``total" K$_{s}$ band; the NASA/IPAC Extragalactic Database).
Second, we subtract  from the image a median background evaluated outside the boundary  and then sum the intensities vertically (perpendicular to the major axis, but within the ellipse) to obtain a 1-D 
 distribution of the net 22~\micron\ flux along the major axis  of the galaxy.
Third, assuming a
 symmetry of the intrinsic flux distribution relative to the center of the galaxy, we take a median average at each off-center data point to minimize potential undesirable inputs from such interlopers  as foreground stars.
 Fourth,  we integrate the resulting distribution to obtain the total 22~\micron\  flux and the effective radius ($R_{W4}$) of the SF galactic disk, which enclose the 90\%  of the 22~\micron\  flux of the galaxy. Finally, we combine the flux and the distance (Table~\ref{t:sam}) to obtain the luminosity  ($\nu L_{22}$) and convert it to  the SFR, using the relation~\citep{jarrett2013}: $ SFR ({M_\odot \, {\rm yr^{-1}}})     = 7.50  \times 10^{-10} \, \nu L_{22} (L_\odot)$. 
 
 The above procedure works for all our sample galaxies, except for M82, the {\sl WISE} W4 intensity of which  is saturated in the central region. We thus replace the SFR of the galaxy with that estimated from the far-IR luminosity as obtained in Paper I.
For the rest of the galaxies, we find that the SFRs estimated from the {\sl WISE} data correlate tightly with
those from the far-IR emission of the galaxies (Paper I), although considerable
differences (by a factor up to a few) between the two estimates do exist for 
individual galaxies with relatively low SFRs ($\lesssim 1 M_\odot~{\rm yr^{-1}}$).
We further estimate the surface SFR as $I_{SFR} = SFR/(\pi R^2_{W4})$.

\begin{table*}
\begin{center}
\begin{minipage}[b]{6.2in}
\caption{Parameters of the sample galaxies}
\begin{tabular}{@{}llccccccccc}
\hline\hline
Galaxy & Galaxy & $d$ & SFR & $R_{W4}$ & $I_{SFR}$ & $D_{25}$ & $L_{K}$ & $M_*$ & $L_{X}$ & $T_{X}$ \\
\# & Name  & Mpc & $M_\odot~\rm yr^{-1}$ & arcmin & $M_\odot~\rm yr^{-1}~kpc^{-2}$ & arcmin & $10^{10}L_{\odot,K}$ & $10^{10}M_\odot$ & $10^{38}\rm~ergs~s^{-1}$ & keV \\
\hline
1 	&	 IC2560 	&	 29.2 	&	 2.60 &	0.53 &	0.0403  	&	 3.5 	&	 $5.30\pm0.14$ 	&	  $1.08\pm0.03$ 	&	 $115.2\pm7.0$ 	&	 -  \\
2 	&	 M82	 	&	 3.5 	&	 7.70 &	1.72 &  0.7870 	        &        11.0   &        $3.57\pm0.05$ 	&	  $1.99\pm0.03$ 	&	 $117.4\pm0.5$ 	&	 $0.611\pm0.003$ \\
3 	&	 NGC0024 	&	 9.1 	&	 0.03 &	1.41 &	0.0007  	&	 6.2 	&	 $0.37\pm0.01$ 	&	  $0.153\pm0.003$ 	&	 $1.7\pm0.3$ 	&	 -  \\
4 	&	 NGC0520 	&	 27.8 	&	 5.29 &	0.47 &	0.1161  	&	 4.1 	&	 $6.37\pm0.11$ 	&	  $3.70\pm0.06$ 	&	 $19.4_{-6.3}^{+3.8}$ 	&	 $0.29_{-0.03}^{+0.05}$ \\
5 	&	 NGC0660 	&	 14.7 	&	 4.30 &	0.66 &	0.1699  	&	 4.6 	&	 $5.05\pm0.07$ 	&	  $2.98\pm0.04$ 	&	 $11.8_{-2.0}^{+3.1}$ 	&	 $0.52_{-0.13}^{+0.11}$ \\
6 	&	 NGC0891 	&	 9.9 	&	 2.04 &	3.00 &	0.0087  	&	 13.0 	&	 $8.53\pm0.13$ 	&	  $4.94\pm0.07$ 	&	 $38.0\pm0.9$ 	&	 $0.34\pm0.01$ \\
7 	&	 NGC1023 	&	 11.6 	&	 0.05 &	2.40 &	0.0003  	&	 7.4 	&	 $8.47\pm0.12$ 	&	  $6.76\pm0.10$ 	&	 $2.9_{-0.7}^{+0.6}$ 	&	 $0.26_{-0.02}^{+0.03}$ \\
8 	&	 NGC1380 	&	 21.2 	&	 0.14 &	1.00 &	0.0012  	&	 4.6 	&	 $16.10\pm0.24$ 	&	  $12.30\pm0.18$ 	&	 $40.1_{-3.8}^{+3.5}$ 	&	 $0.33\pm0.02$ \\
9 	&	 NGC1386 	&	 15.3 	&	 1.00 &	0.40 &	0.0984  	&	 3.6 	&	 $2.80\pm0.04$ 	&	  $1.83\pm0.03$ 	&	 $15.0_{-2.0}^{+1.2}$ 	&	 $0.26\pm0.01$ \\
10 	&	 NGC1482 	&	 19.6 	&	 4.15 &	0.41 &	0.2427  	&	 2.5 	&	 $3.10\pm0.05$ 	&	  $2.29\pm0.04$ 	&	 $75.1_{-6.7}^{+6.2}$ 	&	 $0.38\pm0.03$ \\
11 	&	 NGC1808 	&	 12.3 	&	 7.17 &	0.72 &	0.3449  	&	 5.4 	&	 $6.60\pm0.10$ 	&	  $3.78\pm0.06$ 	&	 $25.8_{-2.9}^{+2.6}$ 	&	 $0.58_{-0.06}^{+0.03}$ \\
12 	&	 NGC2787 	&	 13.0 	&	 0.02 &	0.34 &	0.0044  	&	 3.2 	&	 $4.30\pm0.04$ 	&	  $3.33\pm0.03$ 	&	 $1.8_{-1.3}^{+1.9}$ 	&	 $0.18_{-0.18}^{+0.11}$ \\
13 	&	 NGC2841 	&	 14.1 	&	 0.66 &	2.31 &	0.0023  	&	 6.9 	&	 $14.77\pm0.22$ 	&	  $9.86\pm0.15$ 	&	 $19.8_{-3.2}^{+2.8}$ 	&	 $0.41_{-0.04}^{+0.07}$ \\
14 	&	 NGC3079 	&	 16.5 	&	 2.50 &	1.29 &	0.0206  	&	 8.2 	&	 $6.77\pm0.10$ 	&	  $2.97\pm0.04$ 	&	 $85.9_{-4.9}^{+4.7}$ 	&	 $0.51\pm0.02$ \\
15 	&	 NGC3115 	&	 9.8 	&	 0.09 &	3.72 &	0.0003  	&	 7.1 	&	 $8.66\pm0.13$ 	&	  $6.79\pm0.10$ 	&	 $0.4_{-0.2}^{+0.1}$ 	&	 $0.08_{-0.08}^{+0.04}$ \\
16 	&	 NGC3198 	&	 14.5 	&	 0.71 &	1.83 &	0.0038  	&	 6.5 	&	 $2.73\pm0.05$ 	&	  $1.04\pm0.02$ 	&	 $16.7\pm1.8$ 	&	 -  \\
17 	&	 NGC3384 	&	 11.8 	&	 0.03 &	2.36 &	0.0001  	&	 5.2 	&	 $5.44\pm0.08$ 	&	  $4.11\pm0.06$ 	&	 $12.6\pm2.2$ 	&	 -  \\
18 	&	 NGC3412 	&	 11.5 	&	 0.01 &	0.51 &	0.0012  	&	 4.0 	&	 $2.28\pm0.03$ 	&	  $1.66\pm0.02$ 	&	 $9.8\pm1.1$ 	&	 -  \\
19 	&	 NGC3521 	&	 11.2 	&	 2.37 &	1.95 &	0.0187  	&	 8.3 	&	 $12.17\pm0.18$ 	&	  $7.05\pm0.10$ 	&	 $26.3\pm1.8$ 	&	 $0.36_{-0.02}^{+0.03}$ \\
20 	&	 NGC3556 	&	 10.7 	&	 1.44 &	1.99 &	0.0120  	&	 4.0 	&	 $3.38\pm0.05$ 	&	  $1.61\pm0.03$ 	&	 $13.1_{-2.6}^{+2.3}$ 	&	 $0.33\pm0.02$ \\
21 	&	 NGC3628 	&	 13.1 	&	 2.77 &	2.62 &	0.0088  	&	 11.0 	&	 $12.03\pm0.18$ 	&	  $6.73\pm0.10$ 	&	 $38.7_{-3.3}^{+2.9}$ 	&	 $0.32\pm0.01$ \\
22 	&	 NGC3877 	&	 14.1 	&	 0.60 &	1.32 &	0.0065  	&	 5.4 	&	 $3.22\pm0.05$ 	&	  $1.75\pm0.03$ 	&	 $4.3_{-2.4}^{+2.2}$ 	&	 $0.30_{-0.06}^{+0.05}$ \\
23 	&	 NGC3955 	&	 20.6 	&	 0.87 &	0.45 &	0.0377  	&	 4.1 	&	 $2.90\pm0.06$ 	&	  $1.29\pm0.03$ 	&	 $10.9_{-6.3}^{+3.7}$ 	&	 $0.31_{-0.05}^{+0.29}$ \\
24 	&	 NGC3957 	&	 27.5 	&	 0.11 &	0.41 &	0.0034  	&	 3.2 	&	 $5.46\pm0.08$ 	&	  $1.11\pm0.02$ 	&	 $19.8\pm3.0$ 	&	 -  \\
25 	&	 NGC4013 	&	 18.9 	&	 0.73 &	1.36 &	0.0041  	&	 4.9 	&	 $6.37\pm0.09$ 	&	  $4.49\pm0.07$ 	&	 $23.3\pm2.2$ 	&	 -  \\
26 	&	 NGC4111 	&	 15.0 	&	 0.07 &	0.61 &	0.0031  	&	 1.8 	&	 $4.37\pm0.06$ 	&	  $3.02\pm0.04$ 	&	 $6.7_{-3.8}^{+2.4}$ 	&	 $0.44_{-0.10}^{+0.12}$ \\
27 	&	 NGC4217 	&	 19.5 	&	 1.54 &	1.37 &	0.0081  	&	 5.4 	&	 $6.78\pm0.11$ 	&	  $4.25\pm0.07$ 	&	 $75.9\pm5.9$ 	&	 -  \\
28 	&	 NGC4244 	&	 4.4 	&	 0.02 &	5.17 &	0.0002  	&	 16.2 	&	 $0.234\pm0.005$ 	&	  $0.088\pm0.002$ 	&	 $1.1\pm0.1$ 	&	 -  \\
29 	&	 NGC4251 	&	 19.6 	&	 0.04 &	0.77 &	0.0007  	&	 2.3 	&	 $6.26\pm0.06$ 	&	  $4.23\pm0.04$ 	&	 $23.2\pm4.6$ 	&	 -  \\
30 	&	 NGC4388 	&	 17.1 	&	 2.33 &	0.62 &	0.0771  	&	 5.4 	&	 $3.37\pm0.05$ 	&	  $1.63\pm0.03$ 	&	 $73.1_{-9.2}^{+11.5}$ 	&	 $0.61_{-0.05}^{+0.04}$ \\
31 	&	 NGC4438 	&	 14.4 	&	 0.16 &	0.69 &	0.0062  	&	 9.2 	&	 $5.00\pm0.07$ 	&	  $3.16\pm0.05$ 	&	 $82.9_{-6.2}^{+5.7}$ 	&	 $0.52\pm0.03$ \\
32 	&	 NGC4501 	&	 15.7 	&	 1.85 &	1.65 &	0.0104  	&	 8.6 	&	 $15.67\pm0.23$ 	&	  $8.11\pm0.12$ 	&	 $138.1_{-26.4}^{+25.3}$ 	&	 $0.56_{-0.07}^{+0.05}$ \\
33 	&	 NGC4526 	&	 17.2 	&	 0.36 &	1.07 &	0.0040  	&	 7.0 	&	 $15.41\pm0.23$ 	&	  $11.91\pm0.18$ 	&	 $18.7_{-4.3}^{+3.9}$ 	&	 $0.27_{-0.02}^{+0.04}$ \\
34 	&	 NGC4565 	&	 11.1 	&	 0.70 &	3.71 &	0.0015  	&	 16.7 	&	 $9.39\pm0.13$ 	&	  $5.27\pm0.07$ 	&	 $10.9_{-1.0}^{+1.1}$ 	&	 $0.36_{-0.02}^{+0.04}$ \\
35 	&	 NGC4569 	&	 9.9 	&	 0.50 &	1.20 &	0.0130  	&	 9.1 	&	 $4.13\pm0.06$ 	&	  $2.08\pm0.03$ 	&	 $23.4_{-9.1}^{+4.0}$ 	&	 $0.56\pm0.04$ \\
36 	&	 NGC4594 	&	 9.8 	&	 0.27 &	2.51 &	0.0017  	&	 8.5 	&	 $20.37\pm0.28$ 	&	  $15.47\pm0.21$ 	&	 $39.2_{-2.4}^{+2.0}$ 	&	 $0.60\pm0.01$ \\
37 	&	 NGC4631 	&	 7.6 	&	 1.60 &	3.34 &	0.0093  	&	 14.5 	&	 $2.81\pm0.04$ 	&	  $1.02\pm0.02$ 	&	 $36.4_{-1.4}^{+1.3}$ 	&	 $0.35\pm0.01$ \\
38 	&	 NGC4666 	&	 15.7 	&	 2.59 &	1.23 &	0.0262  	&	 5.0 	&	 $7.65\pm0.11$ 	&	  $4.07\pm0.06$ 	&	 $86.8_{-38.5}^{+14.2}$ 	&	 $0.27_{-0.05}^{+0.04}$ \\
39 	&	 NGC4710 	&	 16.8 	&	 0.39 &	0.54 &	0.0175  	&	 4.4 	&	 $5.23\pm0.08$ 	&	  $3.35\pm0.05$ 	&	 $6.1_{-3.5}^{+0.8}$ 	&	 $0.63_{-0.06}^{+0.10}$ \\
40 	&	 NGC5102 	&	 3.2 	&	 0.004&	2.31 &	0.0003  	&	 9.7 	&	 $0.29\pm0.01$ 	&	  $0.151\pm0.003$ 	&	 $0.6\pm0.1$ 	&	 -  \\
41 	&	 NGC5170 	&	 22.5 	&	 0.41 &	2.62 &	0.0004  	&	 8.0 	&	 $8.07\pm0.14$ 	&	  $4.73\pm0.08$ 	&	 $32.7\pm8.3$ 	&	 -  \\
42 	&	 NGC5253 	&	 4.1 	&	 0.62 &	0.50 &	0.5614  	&	 5.0 	&	 $0.152\pm0.005$ 	&	  $0.049\pm0.001$ 	&	 $1.8\pm0.1$ 	&	 $0.35_{-0.01}^{+0.02}$ \\
43 	&	 NGC5422 	&	 30.9 	&	 0.04 &	0.46 &	0.0007  	&	 2.8 	&	 $5.91\pm0.10$ 	&	  $4.64\pm0.08$ 	&	 $18.5\pm3.4$ 	&	 -  \\
44 	&	 NGC5746 	&	 24.7 	&	 0.85 &	2.12 &	0.0012  	&	 7.2 	&	 $22.17\pm0.33$ 	&	  $14.29\pm0.21$ 	&	 $17.2_{-10.0}^{+6.4}$ 	&	 $0.16_{-0.16}^{+0.12}$ \\
45 	&	 NGC5775 	&	 26.7 	&	 3.92 &	1.34 &	0.0115  	&	 3.7 	&	 $12.00\pm0.15$ 	&	  $6.57\pm0.08$ 	&	 $101.5_{-11.9}^{+10.2}$ 	&	 $0.38_{-0.04}^{+0.05}$ \\
46 	&	 NGC5866 	&	 15.3 	&	 0.18 &	1.10 &	0.0024  	&	 6.3 	&	 $8.37\pm0.12$ 	&	  $5.53\pm0.08$ 	&	 $13.7_{-2.4}^{+2.0}$ 	&	 $0.31_{-0.03}^{+0.04}$ \\
47 	&	 NGC6503 	&	 5.3 	&	 0.08 &	1.21 &	0.0072  	&	 5.9 	&	 $0.67\pm0.01$ 	&	  $0.313\pm0.005$ 	&	 $1.6\pm0.2$ 	&	 $0.42_{-0.06}^{+0.09}$ \\
48 	&	 NGC6764 	&	 26.2 	&	 2.45 &	0.58 &	0.0401  	&	 2.5 	&	 $2.07\pm0.07$ 	&	  $0.99\pm0.03$ 	&	 $187.3_{-78.5}^{+53.9}$ 	&	 $0.75_{-0.11}^{+0.13}$ \\
49 	&	 NGC7090 	&	 6.3 	&	 0.09 &	1.45 &	0.0039  	&	 8.1 	&	 $0.37\pm0.01$ 	&	  $0.148\pm0.003$ 	&	 $0.4_{-0.4}^{+0.3}$ 	&	 $0.44_{-0.14}^{+0.13}$ \\
50 	&	 NGC7457 	&	 13.2 	&	 0.02 &	1.66 &	0.0002  	&	 4.0 	&	 $1.76\pm0.04$ 	&	  $1.22\pm0.03$ 	&	 $5.0\pm1.3$ 	&	 -  \\
51 	&	 NGC7582 	&	 23.0 	&	10.80 &	0.83 &	0.1119  	&	 7.0 	&	 $12.34\pm0.19$ 	&	  $6.77\pm0.11$ 	&	 $102.3_{-19.3}^{+17.1}$ 	&	 $0.67_{-0.07}^{+0.08}$ \\
52 	&	 NGC7814 	&	 18.1 	&	 0.14 &	1.54 &	0.0007  	&	 4.4 	&	 $9.33\pm0.14$ 	&	  $7.02\pm0.10$ 	&	 $28.9\pm4.8$ 	&	 -  \\ 
\hline
\end{tabular}
\medskip
Two parameters are obtained from the {\sl WISE}
band 4 (centered at 22\micron) data: SFR and the 90\% light-enclosed radius
($R_{W4}$), which together give the inferred surface SFR ($I_{SFR}$; see the text for details). Other parameters are directly quoted from \citet{Li13a}, including the redshift-independent distance to each of the galaxies ($d$), the B-band diameter of the projected major axis at the isophotal level $25\rm~mag~arcsec^{-2}$ ($D_{25}$),  the \emph{2MASS} K-band luminosity ($L_{K}$), the stellar mass estimated from $L_{K}$ and a color-dependent mass-to-light ratio ($M_*$), as well as the coronal 0.5-2~keV luminosity ($L_{X}$) and temperature ($T_{X}$).
\label{t:sam}
\end{minipage}
\end{center}
\end{table*}
 
Table~\ref{t:sam} presents the SFR and the SF disk size estimated from the {\sl WISE} data
for each of our 52 sample galaxies, as well as their other relevant parameters used in the present study. We have removed NGC~4342, which was included in 
our original sample (Paper I) because its coronal X-ray emission is largely affected by the ICM ~\citep{Bogdan12}.
Paper I details the measurements of these other parameters, which include the 0.5-2~keV luminosity of the galactic corona ($ L_X$),
its characteristic temperature ($T_{X}$),
the K-band luminosity ($L_K$), and the stellar mass ($ M_*$) for each of our sample galaxies. These parameters are also listed in Table~\ref{t:sam}.
In particular, $ L_X$ is measured within a vertical distance of five times the exponential scale height of the diffuse X-ray intensity profile, after quantitatively removing the contributions from both resolved and unresolved stellar sources and correcting for the absorption due to the foreground parts of the galactic disk. 

We use subsamples defined in Paper~I to compare properties of 
different types of galaxies. Briefly, we use the local galaxy number density ($\rho$; \citealt{Tully88}) to characterize the galaxy environment and define galaxies with $\rho\leq0.6$ as being in the field, while those with $\rho>0.6$ as being clustered; separate early- and late-type disk galaxies with their morphological type code (TC): those with $\rm TC\leq1.5$ (Sa or S0) are defined as early-type, while others are late-type. Different subsamples are plotted as different symbols throughout the paper. 

\section{Analysis and Results}\label{s:res}

Our statistical analysis of the parameters, as summarized in Table~\ref{t:sam}, follows the procedure detailed in Paper~II. Briefly, we first use Spearman's rank order coefficient ($r_{s}$; by definition, $-1 \le r_s \le 1$) to describe the goodness of a correlation and consider $|r_s| \ge 0.6$ or $ 0.3 < |r_s| <  0.6$ as a strong or weak correlation, and $|r_s| \le 0.3$ as no correlation. Second, we characterize a correlation with a simple log-log linear relation, together with the corresponding RMS (around the fitted relation); this latter quantity characterizes the probabilistic nature of the correlation and/or our missing accountability of the multiple parameter dependencies, as well as the measurement uncertainties. Third, we apply the same analysis to bootstrap-with-replacement sampled data to estimate the errors of the measured parameters (e.g., the coefficients of the relation). All errors quoted for the analysis are at the 1$\sigma$ confidence level.

\begin{figure}
\unitlength1.0cm
\centerline{
\includegraphics[width=1.03\linewidth,angle=0]{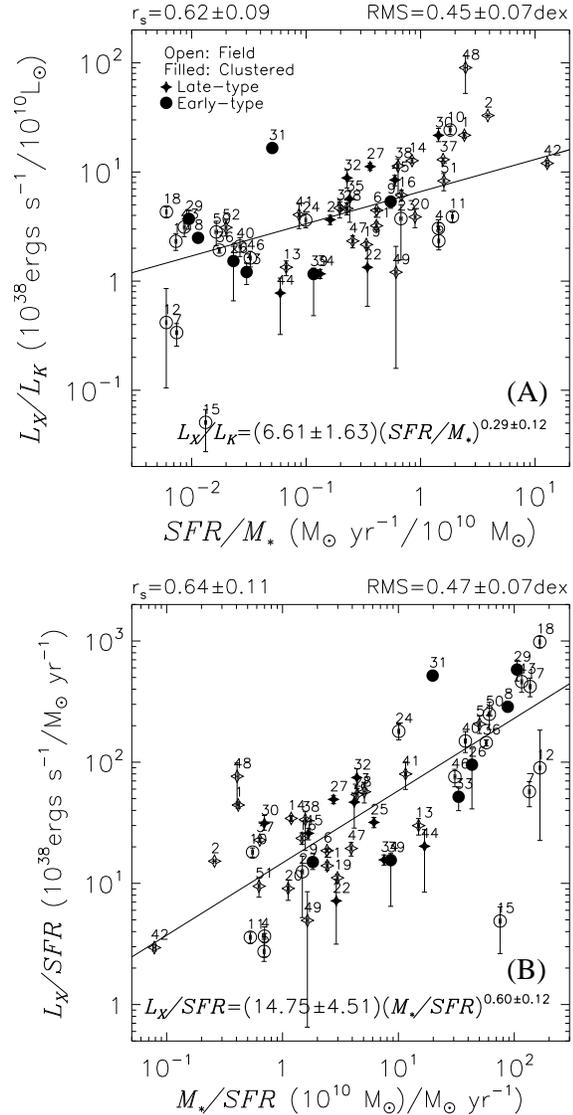}
}
\caption{Correlations among galaxies properties: the 0.5-2~keV luminosity of the galactic corona ($ L_X$), vs. various galaxy properties (see the text for details). In each panel, the solid line shows the best-fitting log-log linear relation; also given are Spearman's rank order coefficient ($r_s$) as well as the RMS scatter around the best-fitting relation. The symbols used in these panels are noted in (A), where the galaxy numbers are listed in the first column of Table~\ref{t:sam}. 
}
\label{f:f1}
\end{figure}

Fig.~\ref{f:f1} shows strong positive correlations of $L_X/L_K$ vs. $SFR/M_*$ 
and $L_X/SFR$ vs. $M_*/SFR$. These two correlations are comparably strong. The fitted log-log linear relations for the correlations are also included in each panel of the figure. Because $L_K\propto M_*$, which is approximately valid (except for a color correction; \citealt{Bell01}; Paper I),  the two
relations, in their power law forms, are equivalent (related by a factor of $M_*/SFR$). Nevertheless, the two panels in the figure help to show the broad (about two orders of magnitude) range spanned by each of the $ L_X/L_K$ and $L_X/SFR$ ratios, separately, and that much of the ranges can be accounted for by the tight correlations of these specific ratios with
$M_*/SFR$. Most importantly, the $ L_X/L_K$ vs. $SFR/M_*$
and $L_X/SFR$ vs. $M_*/SFR$ relations with the power law slopes $\Gamma = 0.29\pm0.12$ and $0.60\pm0.12$ are significantly sub-linear! The larger slope for the latter relation suggests that the galaxy mass is a more important factor than the
specific $SFR$ in determining $L_X$. It is well known
that star-forming galaxies form a sequence in which SFR$/M_*$ declines slowly
 with $M_*$ until it reaches $\sim 10^{11} M_\odot$ (e.g., Fig.~4b in Paper II; \citealt{Brinchmann04,Mitchell14}). Thus in the stellar mass range of our sample
galaxies, $SFR/M_*$, and hence $ L_X/L_K$ (Fig.~\ref{f:f1}), 
hardly depend on $M_*$. As a result, $L_X$ is nearly linearly proportional 
to $L_K$, hence to the SFR. With all these in consideration, we may then 
conclude that the apparent linear correlation between 
$ L_X$ and SFR is largely, although not completely, due to their correlation with the galaxy mass. 

\begin{figure}
\unitlength1.0cm
\centerline{
\includegraphics[width=1.03\linewidth,angle=0]{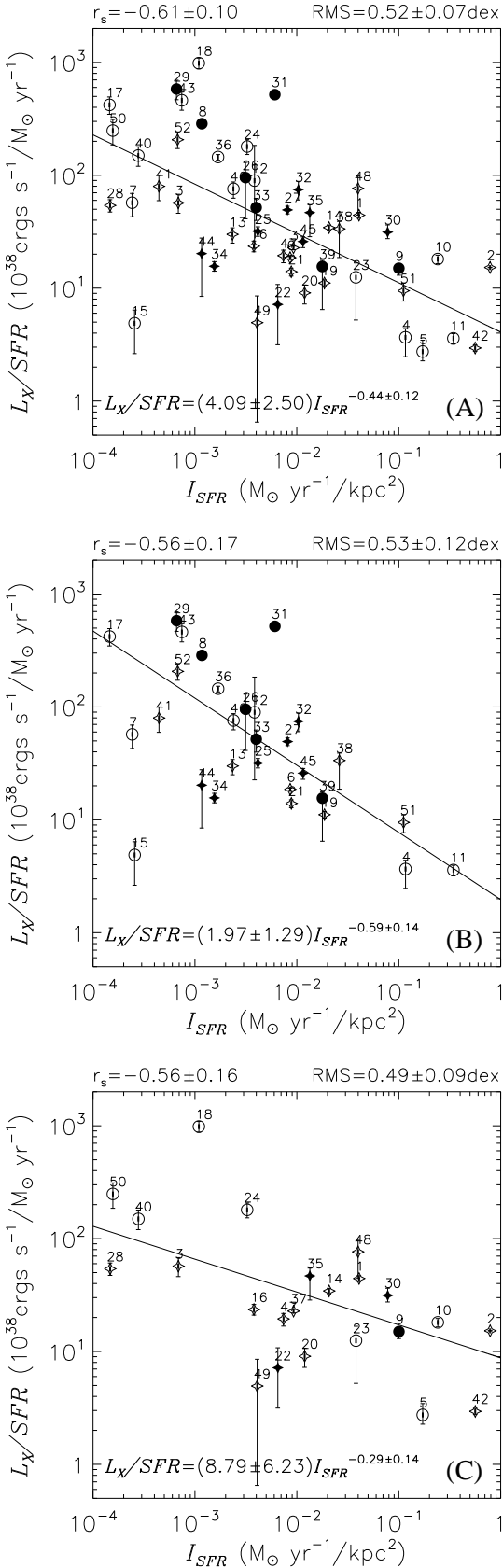}
}
\caption{Anti-correlation between $L_X/$SFR and $I_{SFR}$:
 for the entire sample of galaxies (A), as well as for 
 ones with $M_{*}$ greater (B) or less (C) than 
$3 \times 10^{10} M_\odot$  (which is comparable to the total stellar mass of the Milky Way). The rest is the same as in Fig.~\ref{f:f1}. 
}
\label{f:f2}
\end{figure}

To explore the dependence of the specific X-ray luminosity 
on the SFR further, we present the 
relation of $L_X/$SFR vs. $ I_{SFR}$ in Fig.~\ref{f:f2}. 
The fact that these two quantities are
anti-correlated shows that the efficiency of the coronal emission tends
to decrease with the increasing average strength of the SF in a galactic disk. 
The relation also indicates that $ L_X$ is proportional to the SFR only to a power law of $\Gamma \sim 0.56$ (e.g., for the same galaxy and hence no 
dependence on the mass), if the SF galactic disk area of a galaxy does not change significantly! Indeed, our sample in general shows little correlation between $R_{W4}/D_{25}$ and $SFR/M_*$, except for the two outstanding nuclear starburst galaxies, M82 and NGC5253, each of which has the SFR about an order of magnitude greater than the sample average and is a factor of a few more compact. 
Furthermore, Figs.~\ref{f:f2} B and C suggest that the decrease of 
the coronal emission efficiency
is more profound in massive galaxies than in relatively low-mass ones, as
indicated by the power law slopes of the fits. 
Admittedly, the scatter of the data points is large, especially for the low-mass 
galaxy sample. Therefore, the trend needs to be confirmed with improved
measurements.

\begin{figure}
\unitlength1.0cm
\centerline{
\includegraphics[width=1.03\linewidth,angle=0]{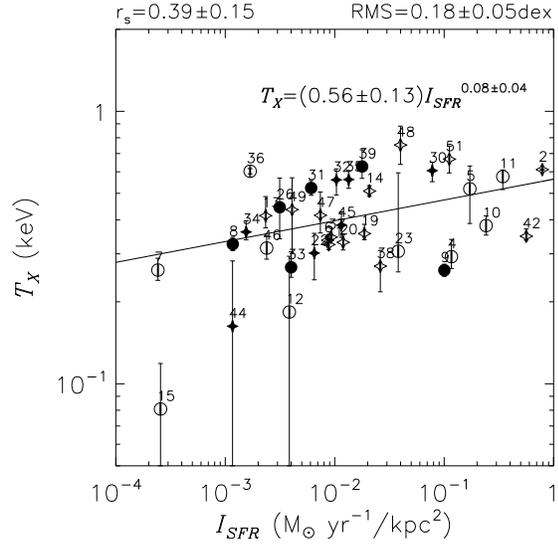}
}
\caption{Correlation between the characteristic temperatures of the
galactic coronae ($ T_X$) and $ I_{SFR}$ for galaxies whose coronal temperature can be well constrained \citep{Li13a}. NGC~3115 is plotted but is removed from fitting the relation and calculating the $r_s$ and RMS, because the diffuse X-ray emission from this galaxy is very weak and the temperature determination is largely affected by the subtraction of the faint stellar component.
The rest is the same as in Fig.~\ref{f:f1}. 
}
\label{f:f3}
\end{figure}

In addition, Fig.~\ref{f:f3} shows only a moderately significant positive correlation of $ T_X$ vs. $I_{SFR}$. This, together with the above strong dependence of the X-ray efficiency on $I_{SFR}$, provides important clues about the origin of the coronal emission.

\section{Discussion}\label{s:dis}

Compared with existing similar statistical studies of the $L_X$-related relations  for disk galaxies (e.g., Paper II;  \citealt{Strickland04b,Tullmann06}),  a key improvement of the present work is the use of the specific parameters. This improvement not only makes the results more reliable and physical (because of the removal of any distance-dependence), but often significantly 
tightens the correlations between various parameters considered here, apparently due to (linearly) accounting for the dependence on the third parameter via the specific scaling. For example, the correlation between $ L_X$ and  $ M_{*}$ (without being scaled) is $r_s=0.29\pm0.15$ (Fig. 1B in Paper II), in contrast to $r_s=0.64\pm0.11$ for  $L_X/$SFR vs. $M_*/$SFR (Fig.~\ref{f:f1}A). In the following, we explore the implications of our newly established correlations and their characteristic relations for understanding the origin of the X-ray emission and the interplay of the stellar feedback with
the galactic environment. 
 
 The origin of the X-ray emission is intimately related to the nature of  galactic coronal gas and its dynamics, which may in turn depend on the circumstance of a galaxy.
 In and around an active star-forming galaxy, especially a low-mass one, coronal gas is expected from the heating of the interstellar medium (ISM) via stellar feedback (primarily core collapsed SNe and fast stellar winds of massive stars). Because such stars are formed in OB associations, this feedback is concerted in both time and space, producing hot superbubbles, which, if energetic enough, can  blow out from cool gas disks of galaxies (e.g., ~\citealt{MacLow88,MacLow99,Melioli13,Keller14}). We have shown above that 
the lower $I_{SFR}$ is, the higher the X-ray emission efficiency becomes! This can naturally be  explained by increasing {\sl specific} mass-loading of the
outflow with decreasing $I_{SFR}$ (see further discussion later), consistent with a lower average temperature  of the coronal gas in a lower SFR galaxy 
(\S~\ref{s:res}). The higher density and lower temperature lead to the
enhanced cooling rate and slow bulk motion of the coronal gas. As a result, the outflow tends 
to form only a localized galactic fountain. 
Here we have assumed that $ L_X$ traces the emission from coronal gas outflow
itself. This assumption is reasonable for our highly-inclined disk galaxies, especially those with normal or low $I_{SFR}$, for which no significant galactic superwinds are expected. Soft X-ray emission from the hot ISM within the thin cool gas disk of such a galaxy is expected to be largely absorbed  and should contribute little to the observed $L_X$. 

Our fitted  relation (Fig.~\ref{f:f2}A),
\begin{equation}
L_X/SFR\propto I_{SFR}^{-0.44},
\label{e:obs}
\end{equation}
provides clues about the outflow process, at least for a galaxy with high $I_{SFR}$. 
We use a parameter $\gamma$ to characterize the effective hot ISM energy fraction
 that ends up in an observed galactic corona. Apparently,
$\gamma$ is proportional to the fraction of OB associations that are energetic
enough to produce blowout superbubbles from the disk. This latter fraction likely depends on the mode of SF  (quiescent vs. 
galaxy merger-triggered) and should increase with increasing $I_{SFR}$~(e.g., \citealt{oey98,Adamo15}; but see also~\citealt{Chandar15}). 
According to the canonical galactic superwind model~\citep{Chevalier85}, which
should be a reasonable approximation for the hot gas outflow in the vicinity of
a galactic disk, one expects
\begin{equation}
L_X/SFR \propto SFR \gamma\beta^{3}/(\alpha R_*), 
\label{e:mod}
\end{equation}
where $\beta = \dot{M}_{hot}/SFR$ is the mass-loading efficiency and $\alpha$ is the part 
of the feedback energy thermalized into the hot gas (determining its
temperature; e.g., ~\citealt{Strickland09,Zhang15}). 
In a superbubble blowout model, $\alpha$ is also a constant, while 
the size of the outflow region $R_*$ is 
proportional to the scale height of the cool gas disk, which may be reasonably 
assumed to be independent of the SFR (e.g., \citealt{MacLow88,MacLow99}). In any case,
the dependencies of the above relation on both $\alpha$ and $R_*$  are weak. Further neglecting a potential weak dependence of the SF disk size on the SFR (\S~\ref{s:res}), 
the comparison of Eqs.~\ref{e:obs} and \ref{e:mod} then gives  
\begin{equation}
\beta \propto \gamma^{-1/3} I_{SFR}^{-0.5}.
\end{equation}
  Clearly, the mass-loading efficiency needs to decrease sharply 
with $I_{SFR}$ with  $\Gamma \lesssim -0.5$ because $\gamma$ is expected to increase with $I_{SFR}$. This decrease may be expected qualitatively as a result of the increasing porosity of  a galactic gaseous disk with $I_{SFR}$~\citep[e.g.,][]{Clarke02}.

When the mass-loading efficiency decreases with $I_{SFR}$ of a galaxy, the specific energy (per gas mass) or temperature of the coronal gas should increase. While this trend is qualitatively consistent  with the positive correlation of $T_X$ vs. $ I_{SFR}$, the fitted slope of the relation ($\Gamma \sim  0.08\pm0.04$) is far smaller than the expected value in a superwind solution 
($\sim 0.7$; ~\citealt{Zhang15}). This indicates that the temperature measurement is biased by
the soft X-ray contribution from a cooler component, most likely at the interface between the hot and cool gases (e.g., via processes such as charge exchange, turbulent mixing, and thermal conduction, as well as the resultant enhanced radiative cooling; e.g.,~\citealt{Liu11,Liu12,Li11,Zhang14}). The entrainment of cool gas is indeed expected in outflows driven by massive SF, as demonstrated in many simulations (e.g.,~\citealt{Strickland00,Marcolini05,Melioli13}), and is apparent in many observations (e.g., ~\citealt{Strickland02,Hoopes05,Heckman15}). In particular, 
\citet{Heckman15} have shown that the velocity of the warm ionized phase, 
which should be directly related to
the bulk motion of the outflow (or its volume-filling gas temperature), correlates strongly 
with both SFR and $I_{SFR}$.
The X-ray emission at  the interface tends to be substantially softer than that from the volume-filling coronal gas  in the outflow itself, which can greatly bias the temperature measurement via X-ray spectral fitting  (e.g., Zhang et al. 2014) and may be responsible for the weak dependence of $T_X$ on $ I_{SFR}$. Therefore, $T_X$ may not be a true representative temperature of the volume-filling coronal gas in starburst galaxies such as M82. Indeed, these inferences are consistent with those obtained
from the detailed modeling of the X-ray emission from M82~\citep{Strickland09}. 
 
 What might be the fate of the coronal gas outflows from active star-forming galaxies? With a low mass-loading efficiency, the coronal gas can hardly lose energy, radiatively, which explains the low X-ray emission efficiency of the galactic coronae in general ($\lesssim 1\%$; Paper II). A considerable fraction of the kinetic energy in the outflows may be consumed in entraining the cool gas (e.g., ~\citealt{Marcolini05}),
in accelerating cosmic-rays and in generating/enhancing magnetic field and turbulence (e.g., ~\citealt{Everett08,Socrates08,salem14,girichidis15,Li15b}). The remaining energy in the outflows may be released at large distances from the galactic disks. The radiation may not even be in the observable X-ray band. The outflowing hot gas needs to overcome the gravitational potential, undergo adiabatic cooling, and probably gradually mix with entrained and/or pre-existing cooler gas in extended galactic halos or beyond (e.g., ~\citealt{Thompson15}). The bulk of the energy in the cooled metal-rich gas may end up in the extreme UV to very soft X-ray range, nearly unobservable because of severe interstellar absorption  in our Galaxy. This ionizing radiation, if accounting for a significant fraction of the stellar feedback energy (e.g., 25\%), could then be as intense as that from stars and be stronger than the extragalactic background in the vicinity of a galaxy (e.g., within 100~kpc galactocentric radius; \citealt{Fox05}). Both the radiative and mechanical heating may be responsible for such highly-ionized species as Ne VIII and O VI, as observed in UV absorption line observations~\citep[e.g.,][]{Tripp11}.
 Finally, the wide redistribution of the stellar feedback energy around galaxies and in a broad energy range may also be responsible for the X-ray faintness of spiral-dominated groups, in which the gravity does not play as a dominant role
as in massive clusters of galaxies~\citep[e.g.,][]{O'Sullivan14}. 

The circumstance in and around a massive disk galaxy (e.g., $M_{*} \gtrsim 3 \times 10^{10} M_{\odot}$) can be quite different. Such a galaxy is expected to have a very extended hot gaseous halo, because of the gravitational heating (due to shocks at the virial radius and subsequent compression as part of the IGM accretion process; e.g.,~\citealt{Keres05,Dekel06,Mitra15}). When the stellar feedback-driven outflow meets the accretion flow, multiple effects on the X-ray radiation efficiency may be expected (e.g., ~\citealt{Tang09,Sarkar15}). On one hand, the outflow is likely terminated in the vicinity of the galactic disk by the presence of the hot gaseous halo (assuming that its density is not too low). This termination can naturally re-thermalize the outflow and potentially enhance the coronal emission. Furthermore, the outflows are also likely of multiple phases. Entrained cool materials can, in principle, mix with the ambient hot gas, triggering its cooling and feeding of the SF in the galactic disk (e.g., ~\citealt{Fraternali13}). Enhanced soft X-ray emission may then be  expected. On the other hand, the outflow inflates the halo gas, as well as heats it via forward shocks or sonic waves, effectively reducing its emission efficiency. The heating may also result from the mixing of the outflow with the halo gas in a very convective or turbulent environment. Indeed, no significant amount of very hot gas ($\gtrsim 10^{7}$~K) has generally been observed, which would otherwise be expected from the re-thermalization of the outflow without quick mixing with entrained or pre-existing halo gas. The anti-correlation between $L_{X}/SFR$ and 
$I_{SFR}$ indicates that the net effect of the stellar feedback is the reduction of the coronal emission efficiency. Or in other words, the gaseous halo gains
energy from the stellar feedback. In fact, without the feedback, cooling flows would be expected around galaxies disks, which are not generally observed (e.g., ~\citealt{Tang09}). Therefore, we may conclude that the rarefying effect of the halo gas due to the feedback appears to be more important than the potential X-ray emission enhancement effects, at least for our sample galaxies. 

\section{Summary and Future Work}\label{s:sum}

We have examined the correlations of the specific diffuse 0.5-2~keV
luminosities of the galactic coronae  (per SFR or stellar mass) with other distance-independent 
galaxy parameters for a sample of 52 \chandra-observed nearby highly-inclined disk galaxies.
In particular, we have used the {\sl WISE} 22\micron\ data of these galaxies to estimate their SFRs and SF disk sizes, which enables a correlation analysis of the luminosities with the surface SFRs. We have further explored the implications of our results on the origin and energetics of the coronal emission and on the interplay of the stellar feedback with the galactic environment. Our findings are as follows: 

\begin{itemize}
\item The specific X-ray luminosities 
are strongly correlated with the specific SFRs of the galaxies
in a substantially sub-linear fashion. This sub-linearity is in sharp contrast to the commonly observed linear correlation between $ L_X$ and SFR, which is largely due to the
correlations of these two parameters with galaxy stellar mass.

\item The specific X-ray luminosity decreases with the increasing surface SFR of a galaxy.
This, together with the above sub-linear dependence, strongly indicates that the radiative cooling of the coronae is 
highly inefficient around the galactic disks, especially those 
starburst ones. This effect may be caused by a decreasing mass-loading 
efficiency of coronal gas outflows with the increasing surface SFR.
However, the volume-filling hot gas in these outflows may not be  
represented by the soft X-ray emission, which most likely traces 
their interaction with entrained and/or pre-existing cool gas in the vicinity of the disks.

\item The interplay of such an outflow with its environment may strongly
depend on galaxy mass. For a relatively low-mass galaxy, the outflow may be
dissipated and mixed with ambient gas in regions far away from the galactic 
disk. The feedback energy may be released largely in the extreme-UV to very soft X-ray range,
significantly contributing to the ionizing radiation around the galaxy.
For a massive galaxy, the outflow may be terminated in the vicinity
of the galactic disk, because of the presence of an expected extended hot gaseous
halo. The energy injection from the outflow termination tends to reduce the 
radiative cooling rate of this halo and hence its gas supply to the disk.
\end{itemize}

The present study represents a step forward in exploring the interplay between the stellar feedback and the circumgalactic medium via the dependencies of the observed coronal emission on a very limited set of galaxy parameters. The coronal luminosity  should also depend on such properties as metallicity, rotation, SF history, interaction with companions, and clustering environment of a galaxy (e.g., \citealt{Sarzi13,Kim13}), which are not accounted for here. These dependencies may be responsible for much of the  large RMSs of the data around our fitted relations of  the correlations. A good example of such dependencies is the Virgo-cluster galaxy, NGC4438 (Galaxy \#31 in the above figures). Its unusually enhanced
coronal luminosity is most likely due to the ongoing strong 
ram pressure stripping/mixing and compression by the surrounding hot 
intracluster medium~\citep[e.g.,][]{Ehlert13,Lu11}. The X-ray luminosity
of an outflow should also depend on its exact stage. For example, 
the major outflow from the galactic nuclear region of M82 (Galaxy \#2) 
is apparently still at an early stage;  coronal gas is largely 
confined and is undergoing strong mixing with surrounding cool gas~\citep[e.g.,][]{Zhang14}, which is at least partially responsible for the relatively large specific 
X-ray luminosity of the galaxy. Our sample also does not include 
(ultra)luminous infrared galaxies. Therefore, our results or conclusions 
may not apply to intensive starburst galaxies, especially those undergoing major
mergers.  We have investigated various other potential dependencies of
the luminosities on, for example, the overall size~\citep{Li13a} of 
galactic coronae, and the inclination angle and cool gas content of 
the galactic disks. However, we find that the existing measurements 
of these parameters are too uncertain to allow us to reach any reasonably 
firm conclusions. In any case, we do not see how these potential dependencies 
may qualitatively change our conclusions.
Therefore, much work still needs to be done to understand complicated galactic ecosystems, as well as to confirm and/or tighten the correlations examined here. 

Observationally, it remains a challenge to map out the large-scale hot circumgalactic medium around disk galaxies (on scales beyond $\sim 10$~kpc), probably except for a few very massive ones (e.g., ~\citealt{Anderson16}).  But as demonstrated in the present work, observations of the X-ray emission from the vicinities of galactic disks can provide  useful constraints on the galactic disk/halo interplay.  Ongoing analysis of deep X-ray observations of carefully selected disk galaxies, especially massive ones, will hopefully provide additional insights. Expanding the  size of the galaxy sample, ideally volume- or flux-limited, will allow for many potentially useful subsample analysis within relatively narrow parameter spaces.

Theoretically, simulations of galaxy formation and evolution need to confront the new constraints, not only in a rough comparison with the observed X-ray luminosity range of galactic coronae, but in predicting their dependencies on the specific and surface SFRs  and other galaxy parameters as well. 
Particularly important is the understanding of relevant 
physical processes at the interfaces between cool and hot gases, including
turbulent mixsing, thermal conduction, and 
charge exchange. These processes may significantly
contribute to the observed X-ray emission~\citep[e.g.,][]{Zhang14}, as well 
as determine the mass-loading, dynamics, and cooling of galactic coronae~\citep[e.g.,][]{Thompson15}. Progress has indeed been made recently, including
the implementation of the coupling between the hot and 
cool gases in galaxy simulations
 by \citet{Keller14,Keller15}, using the superbubble model~\citep{MacLow88}. 
However,
the validity of this implementation needs to be further tested; e.g., X-ray 
spectra, as well as luminosities, should be calculated and be compared 
with observations. Such direct confrontation between theories and observations
can then help us to understand the underlying physical processes, which are
so fundamental to the study of stellar feedback in the context of 
galaxy formation and evolution.

\section*{Acknowledgments}
We thank the referee, as well as Shawn Roberts,
for valuable comments that led to an improved presentation of the paper.  QDW is grateful to the hospitality that he received in the School of Astronomy and Space Science, Nanjing University, during his visit as a Yixing Visiting Chair Professor. TF was partially supported by the National Natural Science Foundation of China under grant No.~11273021, and by the Strategic Priority Research Program ``The Emergence of Cosmological Structures" of the Chinese Academy of Sciences, Grant No. XDB09000000.

\input{ms.bbl}
\end{document}